%% file: rad.tex
\documentstyle[psfig]{mn}

\newcommand{\itl}[1] {{\it #1}}

\input epsf
\def\DIRFIGS{figures/}

\newcommand{\cidfig}[6]{
    \protect\centerline{
    \epsfysize=#1\epsffile[#2 #3 #4 #5]{#6}
    }}

\newcommand{\mboxs}[1]{\mbox{{\scriptsize #1}}}

\title[Monthly Notices: Radiative shocks in galaxy formation (I)]
  {Radiative shocks in galaxy formation.\\I: Cooling of a primordial
plasma with no sources of heating.}
\author[M.I. Forcada-Mir\'{o} and S.D.M. White]
  {M.I.~Forcada-Mir\'{o}$^{1,2}$
  and S.D.M.~White$^2$\\
  $^1$Institute of Astronomy, Madingley Road, Cambridge CB3 0HZ,
  England\\
  $^2$Max-Planck-Institut f\"{u}r Astrophysik, 
  Karl-Schwarzshild-Stra\mbox{\ss}e 1, 85748 Garching b.\ M\"{u}nchen, 
  Germany
  }
\date{Accepted ????. Received ????}
\pagerange{\pageref{firstpage}--\pageref{lastpage}}

\begin{document}

\label{firstpage}

\maketitle

\begin{abstract}
We use a 1-D Lagrangian code which follows both a gaseous and a
dark component to study the radiative shocks that appear in the 
evolution of spherical scale-free perturbations in an Einstein-de 
Sitter Universe. The detailed behaviour of the shock depends on
whether the radiative cooling is dominated by bremsstrahlung
or line cooling. Bremsstrahlung is the main energy loss  mechanism for
systems with circular velocity $V_{c} > 100 \; km \; s^{-1}$. In this
case, we can reproduce the kinematics of the shock and of the cooling 
wave to a high degree of accuracy with a simple analytic model. When
line cooling dominates, the shock front can be unstable to
oscillations. The period and amplitude of the oscillations increase 
in time as the universe expands. For such systems, our analytic model
provides only a rough estimate of the mean evolution. We believe that 
now we fully understand the effect radiative cooling has in the shocks
that appear in spherical models for galaxy formation.
\end{abstract}


\section{Introduction}

\input{sec1rad}

\section{Model and numerical method}

\input{sec2rad}

\section{Tests of numerics}

\input{sec3rad}

\section{Collapse simulations}

\input{sec4rad}

\section{Theoretical model}

\input{sec5rad}

\section{Conclusions}

\input{sec6rad}

\section*{ACKNOWLEDGMENT}

M.I. Forcada I Mir\'{o} would like to thank the Fundaci\'{o}n Rich and the 
Max-Planck Gesellschaft for the economical support. He is thankful as well 
to Amancio Fria\c{c}a, Zoltan Haimann, Thomas Plewa and Martin Rees for 
many helpful discussions.

\include{bibrad}
\include{tab1rad}

\include{tab2rad}

\include{tab3rad}

\include{fig01rad}

\include{fig02rad}

\include{fig03rad}

\include{fig04rad}

\include{fig05rad}

\include{fig06rad}

\include{fig07rad}

\include{fig08rad}

\include{fig09rad}

\include{fig10rad}

\include{fig11rad}

\include{fig12rad}

\include{fig13rad}

\end{document}

%% file: sec1rad.tex
The origin of galaxies is still an open question in modern
cosmology. The standard picture of structure formation 
supposes that the nonlinear objects we observe
today grew through gravitational instability from a spectrum of 
linear primordial fluctuations \cite{pe1}. The cosmic microwave 
background radiation (CMB) can severely constrain such structure 
formation models \cite{wss}, but observations with high
angular resolution are required to put direct constraints on
scales approaching those relevant to galaxy formation. Cosmogonies dominated by cold 
dark matter (CDM) are the current standard, but there are few
unambiguous tests of their predictions so far on galactic scales 
(Blumenthal \itl{et al.} 1984; Davis \itl{et al.} 1985; Ostriker 1993;
Ostriker and Steinhardt 1995). 

In a cold dark matter scenario structure forms hierarchically, which
means that small objects form first and merge to form bigger ones. As
the gas falls into the potential well of the forming halo it passes
through a shock in which its bulk kinetic energy is converted into
thermal energy. In the case when the gas can cool efficiently it will
flow further in, getting very dense. This cold and dense gas in the
innermost parts of halos provides suitable raw material for star
formation. The important role of radiative cooling in structure
formation was first pointed out independently by Binney
\shortcite{bi1}, Rees \& Ostriker \shortcite{ro} and Silk
\shortcite{si}, and the way in which it leads to galaxy formation
at the centres of dark matter halos was first described and modelled
by White \& Rees (1978). If we want to understand the formation of
galaxies it is essential to study how the gas cools and settles
in the central regions of dark matter halos.

Rapid improvements in computer technology have played an important
role in the theory of structure formation, since numerical simulations
can provide detailed quantitative predictions to be compared to real 
data. Until quite recently only the gravitational clustering of the 
collisionless component could be followed (e.g.\ Davis \itl{et al.} 
1985). Over the last few years, however, it has become 
feasible to introduce gas dynamics and cooling/heating processes in 
simulations of galaxy formation (e.g.\ Cen 1992; Katz, Hernquist \& Weinberg 
1992; Navarro \& White 1993; Steinmetz 1996). Such three-dimensional 
simulations require a large amount of CPU-time, and as a result 
exhaustive parameter studies are difficult. Moreover,
numerical artifacts often complicate interpretation of the results, 
and a clear picture of the physics behind the evolution of such
simulations is not easy to build. 

A very useful alternative is provided by semi-analytic models which 
combine an algorithm for predicting the abundance and merging of 
dark-matter halos (or alternatively a high resolution cosmological
N-body simulation of this process) with cooling arguments to prescribe galaxy 
condensation at halo centres, with a simple star formation law, 
and with parametrised models for supernova feedback and for 
spectrophotometric evolution; such models are ideal for parameter 
studies and for direct comparison with observation, and their 
quantitative predictions can agree reasonably well with a wide range 
of data (e.g. White \& Frenk 1991; Kauffmann, White \& Guiderdoni
1993; Cole \itl{et al.} 1994; Kauffmann, Nusser \& Steinmetz 1997).

On the other hand, semi-analytical models are only as good as the
simple models they use for the various physical processes 
involved. One-dimensional numerical simulations are a potential tool 
to improve such models which do not suffer from many of the
inconveniences of their three-dimensional counterparts. However, very 
little has been done in this direction since the pioneering work of 
Larson (1974). The formation and growth of plane-symmetric pancakes 
has been studied by Shapiro \& Struck-Marcell \shortcite{sha1} and
more recently by Anninos \& Norman \shortcite{anno}; Thoul \& Weinberg
studied dissipational spherical collapses onto local density maxima of
a primordial Gaussian density perturbation field \cite{twe1} 
and the effects of a background of ultraviolet radiation 
\cite{twe2}; Haiman et al.\ \shortcite{zol} adapted the code 
built by Thoul and Weinberg to investigate radiative cooling by
molecular hydrogen. The present paper lies within this general
framework and focusses on the study of the radiative shocks that
appear in the evolution of spherical scale-free perturbations in an 
Einstein-de Sitter universe. It is organized as follows: in Section 2
we present our basic model for formation of a galaxy, as well as the 
1D-Lagrangian code we use to study its evolution; in Section 3 we 
test the code against some existing solutions; in Section 4 we present 
the results of our simulations of galaxy formation; finally, in
Section 5 we present physical explanations for the behaviour observed 
in the simulations.

%% file: sec2rad.tex
The observed rotation curves of spiral galaxies  and the orbital
velocities of their satellites seem to indicate that the surrounding 
dark halos have an approximately isothermal profile $\rho \sim r^{-2}$
(e.g.\ Casertano \& Van Gorkom 1991; Zaritsky \& White 1994). An 
appropriate 1-D model for halo formation should build up
such a halo through infall of material from an expanding universe. 
Gott \shortcite{go1} and Gunn \shortcite{gu1} were the first to 
construct a model of this type. They considered accretion onto a
central seed from an otherwise unperturbed universe, and showed 
that the resulting density profile had $\rho \propto r^{-9/4}$ 
for an Einstein-de Sitter background cosmology. By allowing a 
wider variety of scale-free initial perturbations Fillmore \& 
Goldreich \shortcite{fg} showed that a range of power law 
exponents (including $-2$) could be obtained. White \& Zaritsky 
\shortcite{wz} (thereafter WZ) extended these self-similar infall 
models in two ways. They showed how to embed them in an open 
universe and how the infalling particles could be given nonradial 
orbits. In the WZ model the initial specific binding energy of a 
shell is perturbed in the following way:
\begin{equation}
 \delta E(M) = \delta E_{o} { \left( \frac{M}{M_{o}} \right)
}^{\frac{2}{3} - \epsilon} \; ,
\end{equation}
where $M$ is the mass initially inside the shell. A halo with a 
flat circular velocity curve near the centre results for $\epsilon 
= 2/3$. We use this particular case to generate our initial 
conditions.

\subsection{Collisionless component}

In the model of WZ the equations for the evolution of the
collisionless component are:
\begin{equation}
 \frac{d^{2} \! r_{i}}{dt^{2}} = - \frac{G \; M_{i} \;
r_{i}}{{({r_{i}}^{2} + a^{2})}^{\frac{3}{2}}} - \frac{G \; M_{i} 
\; K_{i} \; J_{i} }{{r_{i}}^{2}} + \frac{{J_{i}}^{2}}{{r_{i}}^{3}} 
\; , 
\end{equation}
\begin{equation}
 \frac{dJ_{i}}{dt} = G \; M_{i} \; K_{i} \; \frac{dr_{i}}{dt} \; ,
\end{equation}
\( G \; M_{i} \; K_{i} = \left\{ \begin{array}{ll}
{(-2E_{i})}^{\frac{1}{2}} \! \sqrt{\frac{1-e}{1+e}} & t \leq
{t}_{i} \\ 0 & t > {t}_{i} \end{array} \right.  \; , \)
\hfill (7) \setcounter{equation}{4}
\begin{equation}
 M_{i} = \sum_{r_{j}  <  r_{i}} \! m_{j} \; ,
\end{equation}
where $m_{i}$, $r_i$, $J_i$ and $t_i$ are the masses, radii,
specific angular momenta, and turnround times of the shells, 
$M_{i}$ is the total mass within shell $i$, and $e$, the 
eccentricity of the final orbit, is assumed to be the same 
for all shells (see WZ for more details).

In order to avoid a divergence of the gravitational acceleration 
near the origin the potential is softened in the usual way. By 
choosing an appropriate value of $e$ we also give each shell 
enough angular momentum to prevent it from getting very close 
to the center. In most of our calculations we adopt $e=0.7$ which 
results in almost isotropic velocity dispersions in the virialised 
regions of the halo.

We integrate equations (2-5) by means of the force-average 
method. First of all we predict new positions and velocities 
using the accelerations calculated at the beginning of the 
time-step ${f_{i}}^{n}$:
\begin{equation}
{r_{i}}^{*} = {r_{i}}^{n} + {\Delta}_{DM} t \; {v_{i}}^{n} + 0.5 
\; {\Delta t}_{DM}^{2} \; {f_{i}}^{n} \; ,
\end{equation}
\begin{equation}
{v_{i}}^{*} = {v_{i}}^{n} + {\Delta t}_{DM} \; {f_{i}}^{n} \; .
\end{equation}
Now, with the predicted positions we can calculate accelerations
at the end of the time step ${f_{i}}^{*}$ and use a centred scheme 
to get positions and velocities for the beginning of the next step:
\begin{equation}
{v_{i}}^{n+1} = {v_{i}}^{n} + 0.5 \; {\Delta t}_{DM} \; \left( 
{f_{i}}^{n} + {f_{i}}^{*} \right) \; ,
\end{equation}
\begin{equation}
{r_{i}}^{n+1} = {r_{i}}^{n} + 0.5 \; {\Delta t}_{DM} \; \left( 
{v_{i}}^{n} + {v_{i}}^{n+1} \right) \; .
\end{equation}
Finally we just set ${f_{i}}^{n+1}={f_{i}}^{*}$ and the step is
complete. With this scheme we only have to calculate the forces 
once per timestep. The timestep ${\Delta t}_{DM}$ is taken as:
\begin{equation}
 {\Delta  t}_{DM} = \min_{i} \left( C_{DM} \! \frac{a}{|v_{i}|} 
\right) \; ,
\end{equation}
where $C_{DM} = 0.1$, so that no individual particle moves more 
than $10 \%$ the softening length in a timestep.

\subsection{Gas component}

The initial conditions for the gas are generated in the same way as
for the dark matter but with no angular momentum. We integrate the
usual hydrodynamic equations, which in spherical symmetry can be
written in the Lagrangian form,
\begin{equation}
 \frac{\partial m}{\partial r} = 4 \pi r^{2} \rho \; ,
\end{equation}
\begin{equation}
 \frac{dv}{dt} = - 4 \pi r^{2} \frac{\partial \! P}{\partial m} -
\frac{4 \pi}{r} \frac {\partial ( r^{3} Q)}{\partial m} - \frac{G
M_{t} r}{{(r^{2} + a^{2})}^{\frac{3}{2}}} \; ,
\end{equation}
\begin{equation}
\frac{du}{dt} = - 4 \pi P  \frac{\partial (r^{2} v)}{\partial m} + 
\frac{3}{2} \frac{Q}{\rho} \left( \frac{\partial v}{\partial r} - 
\frac{\partial (r^{2} v)}{\partial r^{3}} \right) + \frac{1}{\rho}
(\Gamma - \Lambda) \; ,
\end{equation}
\( Q = \left\{ \begin{array}{ll} 3 l^{2} \rho \frac{\partial (r^{2}
v)}{\partial r^{3}} \left( \frac{\partial v}{\partial r} -
\frac{\partial (r^{2} v)}{\partial r^{3}} \right) & \mbox{if
$\frac{\partial (r^{2} v)}{\partial r^{3}} < 0$} \\ 0 &
\mbox{otherwise} \end{array} \right. \; , \)
\hfill (13) \setcounter{equation}{13}
\begin{equation}
 P = (\gamma - 1) \rho u \; ,
\end{equation}
where $\rho$, $v$, $P$ and $u$ are the  gas mass density, velocity,
pressure and specific internal energy of the gas, $\Lambda$ the 
radiative cooling rate, $\Gamma$ the photoionization heating rate, 
$\gamma$ the ratio of specific heats, and $M_{t}$ the total mass 
(gas and dark matter) inside the shell at radius $r$. In order to treat
shocks accurately we have implemented the Tensor Artificial Viscosity 
scheme of Tscharnuter \& Winkler \shortcite{tsw} (TAV). The artificial 
viscous pressure $Q$ is only non-zero in regions that are undergoing
compression (i.e.\ the divergence of the velocity field is negative), 
and converts bulk kinetic energy into internal energy over the scale 
height $l$, which is taken to be $l = \lambda \Delta r $, where
$\Delta r $ is the radial thickness of the shell. The parameter 
$\lambda$ has to be big enough to damp out post-shock oscillations 
but small enough to have a good shock resolution. We have found that 
a value $\lambda = \sqrt{6}$ works well and smears the shock over six
zones.

We set up a logarithmic grid in which equations (11-14) are 
discretized in the way prescribed by Benz \shortcite{be1} (for 
details see Forcada-Mir\'{o} 1997). Time integration is performed 
explicitly, except for the radiative processes, for which we switch 
between an explicit and an implicit method (see 2.3). In this way the 
timestep never becomes too small when strong cooling occurs, but
still, the limitations on the timestep for the gas are stronger than 
for the dark matter. Consequently, we allow different timesteps for
the two components. The timestep for the gas is taken as
\begin{equation}
 {\Delta t}_{g} = \min ({\Delta t}_{CFL}, {\Delta t}_{DM} ) \; ,
\end{equation}
where 
\begin{equation}
 {\Delta t}_{CFL} = \min_{i} \left( C_{CFL} \! \frac{r_{i} -
r_{i-1}}{| v_{i} - v_{i-1} | + \sqrt{\gamma (\gamma - 1) 
{u}_{i}}} \right) \; ,
\end{equation}
and ${\Delta t}_{DM}$ is given in equation (10). In the simulations 
we use $C_{CFL} = 0.25$. The way we couple the two components is by 
advancing the dark matter particles a timestep ${\Delta t}_{DM}$, and 
then repeatedly advancing the gas until a further step ${\Delta
t}_{g}$ would take the gas ahead of the dark matter. At this point we 
just take the timestep required to synchronize the two components. The
contribution of the dark matter to the total mass $M_{t}$ inside each 
gas shell is linearly interpolated in time. We use a force-average 
scheme similar to equations (6) to (9) to integrate the gas equations,
but in this case the final advanced quantities are used to re-evaluate
the accelerations at the end of each step before starting the next
one. Forces are thus computed twice per time step.

\subsection{The cooling}

We consider a primordial plasma (i.e. hydrogen and helium only)
with a helium mass fraction $Y = 0.24$. The integrations are started
at redshift $z_{i} = 500$, just after recombination. The temperature
of the gas is assumed to be uniform and equal to the temperature of
the CMB, i.e. $T = 2.726 (1+z_i)$~K \cite{mat}. The initial
fractional abundances relative to the total number density of hydrogen
are taken to be:
\begin{equation}
x_{{H}^{+}} = x_{e} = 1.2 \times 10^{-5} \frac{{{\Omega}_{o}}^{1/2}}
{{\left( h{\Omega}_{B} \right)}^{-1}} \; ,
\end{equation}
\begin{equation}
x_{{He}^{+}} = x_{{He}^{++}} = 0 \; .
\end{equation}
Equation (17) is the fit to the residual ionization fraction of
hydrogen by Peebles \shortcite{pe4}. Helium would have recombined at
an earlier epoch, so we can assume it neutral \cite{sgb}. In all the 
simulations presented in this paper we took $h = 0.5$ and
${\Omega}_{o} = 1$.

We use the ionization and recombination coefficients, and the cooling
and heating rates revised by Abel (private comunication). In order to 
reduce the computational time we interpolate the coefficients and
rates out of a table constructed beforehand. One has to bear in mind 
that there are many physical processes that can affect the emissivity 
of a hot primordial plasma in a drastic way (e.g.\ the presence of 
ionizing radiation, thermal conductivity by electrons, molecular
hydrogen formation) and it is beyond the scope of this paper to study 
all their effects. 

We have assumed that electrons and protons have the same temperature. 
At the shock front it is the ions, and not the electrons, that are 
initially heated; the electrons are heated later by some mechanism 
which is at least as fast as Coulomb interactions with the positive 
ions \cite{sham}. The Coulomb equilibration time is given by
\begin{equation}
 t_{e-i} n_{H} \approx 9.5 \frac{1.-Y}{1.-0.75Y} {\left[
\frac{T}{{10}^{5} \; \mbox{K}} \right]}^{ \frac{3}{2} } 
\mbox{yr cm}^{-3} \; ,
\end{equation} 
\cite{spi}, and the cooling time, defined as
\begin{equation}
 t_{cool} = \frac{(3/2) n kT_{vir}}{{n_{H}}^{2} \Lambda (T_{vir})} \; ,
\end{equation}
can be written more conveniently as
\( \begin{array}{lcl} t_{cool} n_{H} & \approx & \frac{6.6 \times
10^{4}}{(1-Y) \mu} {\left[ \frac{\Lambda (T)}{10^{-23} \; \mbox{erg 
cm}^{-3} \; \mbox{s}^{-1}} \right]}^{-1} \\
\\
 & & \times \left[ \frac{T}{10^{5} \; \mbox{\scriptsize{K}}} 
\right] \; \mbox{yr cm}^{-3} \end{array} \; , \) \hfill (22) 
\linebreak \linebreak
\setcounter{equation}{22}
where $\mu$ is the mean molecular weight and $Y$ is the primordial
helium fraction by mass. In Figure 1a one can see that the Coulomb 
equilibration time is smaller than the cooling time up to very high 
temperatures, of the order of $10^{8}$~K. It is thus a reasonable
assumption to adopt a single post-shock temperature for
electrons and ions.

Whenever a mostly neutral plasma gets shock-heated up to the virial 
temperature $T_{vir}$, the ionization time, \hfill \linebreak \(
t_{ion} = {(n_{e} \alpha)}^{-1} \), can be longer than the cooling
time ($n_{e}$ is the electron number density and $\alpha$ is the
collisional ionization rate). The implication is that ionization of
the plasma lags behind shock-heating, and line cooling can be enhanced
by the higher neutral fractions. As one can see in Figure 1a, this is
the case up to temperatures of the order of $10^{6}$~K (i.e.\ $V_{c}
\sim 160$ km~s$^{-1}$). On the other hand, the UV flux radiated  
by the hot shocked plasma can preionize hydrogen and helium
upstream of the shock provided $V_{s} > 110$ km~s$^{-1}$ \cite{shm},
where $V_{s}$ is the shock velocity in the rest frame of the shock.
In our systems $V_{s} \sim 1.2 \; V_{c}$, so we expect the plasma to 
be preionized for halos with $V_{c} > 100$ km~s$^{-1}$. We do not take
this effect into account in this paper.

In higher circular velocity halos, the postshock heated gas quickly 
relaxes to the ionization equilibrium abundances. As the plasma
reaches higher densities the cooling time decreases and eventually the
gas begins to cool efficiently, while maintaining a higher ionized 
fraction. The reason is that the recombination time-scales, \hfill 
\linebreak \( t_{rec} = {(n_{e} \beta)}^{-1} \), are much longer than 
the cooling time-scale ($\beta$ is the recombination rate). In Figure 
1b we can see that if the gas gets heated above $10^{6}$~K it will 
cool out of ionization equilibrium. 

Because of the various possibilities, we do not assume equilibrium 
abundances for the different ions, but rather solve the full rate 
equations,
\begin{equation}
 \begin{array}{lcl}
\frac{\textstyle 1}{{\textstyle x}_{e}} 
\frac{\mbox{d} {\textstyle x}_{{\mboxs{H}}^{+}}}{\mbox{d}
\tilde{\textstyle t}} & =  & \frac{\textstyle 1}{{\textstyle \cal C}}
\; {\tilde{\rho}}_{g}
\left\{ {\beta}_{{\mboxs{H}}^{o}} x_{{\mboxs{H}}^{o}} - 
{\alpha}_{{\mboxs{H}}^{+}} x_{{\mboxs{H}}^{+}} \right\} 
 \end{array} \; ,
\end{equation}
\( \begin{array}{lcl} \frac{\textstyle 1}{{\textstyle x}_{e}} 
\frac{\mbox{d} {\textstyle x}_{{\mboxs{He}}^{+}}}{\mbox{d} 
\tilde{\textstyle t}} & = & \frac{\textstyle1}{{\textstyle \cal C}} \;
 {\tilde{\rho}}_{g} \left\{ {\beta}_{{\mboxs{He}}^{o}} 
x_{{\mboxs{He}}^{o}} \; + \; {\alpha}_{{\mboxs{He}}^{++}} 
x_{{\mboxs{He}}^{++}} \right. \\ 
 & & \left. - \; \left( {\beta}_{{\mboxs{He}}^{+}} + 
{\alpha}_{{\mboxs{He}}^{+}} \right) x_{{\mboxs{He}}^{+}} \right\} 
\end{array} \; , \) \hfill (24) \setcounter{equation}{24}
\begin{equation}
 \begin{array}{lcl}
\frac{\textstyle 1}{{\textstyle x}_{e}} \frac{\mbox{d} 
{\textstyle x}_{{\mboxs{He}}^{++}}}{\mbox{d} \tilde{\textstyle t}} 
& = & \frac{\textstyle 1}{{\textstyle \cal C}} \; {\tilde{\rho}}_{g}
\left\{ {\beta}_{{\mboxs{He}}^{+}} x_{{\mboxs{He}}^{+}} - 
{\alpha}_{{\mboxs{He}}^{++}} x_{{\mboxs{He}}^{++}} \right\} \; ,
 \end{array}
\end{equation}
where $x_{i}$ is the fractional abundance of the species $i$ relative
to hydrogen, and ${\beta}_{i}$ (${\alpha}_{i}$) are the ionization
(recombination) rates in units of cm$^{3}$ s$^{-1}$. The time $\tilde
t$ is given in the units,
\begin{equation}
{\cal T} = {H_{o}}^{-1} = 9.78 \; h^{-1} \; \mbox{Gyr} \; ,
\end{equation}
the density ${\tilde{\rho}}_{g}$ in the units,
\begin{equation}
{\cal D} = \frac{3 {H_{o}}^{2}}{4 \pi G} = 3.753 \times 10^{-29} h^{2}
\; {\mbox{gr cm}}^{-3} \; ,
\end{equation}
and the factor ${\cal C}$ has the units of a reaction rate,
\begin{equation}
{\cal C} = 1.429 \times 10^{-13} (1 - Y_{\mboxs{P}})^{-1} \; h \; \; 
{\mbox{cm}}^{3} {\mbox{s}}^{-1} \; .
\end{equation}
Equations (23)-(25) are written under the assumptions that the coronal
limit is valid and the total number density of hydrogen varies on a
much longer time scale than the ionization state of the plasma. It
might happen that the problem is stiff, and then, cannot be handled
efficiently by many integration schemes. We use the subroutine LSODAR,
coded by L. Petzold and A. Hindmarsh to deal with stiff ordinary
differential equations \cite{phi}. 

If cooling/heating terms in the energy equation (equation 13) are
treated explicitly, the gas timestep has to be limited as well by the
cooling time (equation 20). However, the timestep will become to small
whenever strong cooling occurs. To prevent this from happening, the
code uses an operator splitting scheme \cite{press}. In a real case
the microphysical processes operate continuously while the flow is
evolving, but there is no way of maintaining this simultaneity without
paying a big price in the total computing time. The temperature at the
end of a timestep $T_{i}^{n+1}$ is the solution to the nonlinear
algebraic equation,
\begin{equation}
 \begin{array}{lcl}
 \frac{T_{i}^{n+1}}{T_{\mboxs{vir}}} & = & 2 (\gamma - 1) \; 
 \frac{{\mu}_{i}^{n+1}({\tilde{\rho}}_{g_{i}}^{{ \; \: }^{n+1}}, 
 T_{i}^{n+1})}{{\mu}_{o}} \; \left[ 
 ({\tilde{u}}_{i}^{n+1})_{\mboxs{ad}} \; + \right. \\
  & & \left. \frac{1}{{\cal L}} \; {\tilde{\rho}}_{g_{i}}^{{ 
 \; \: }^{n+1}} \; S({\tilde{\rho}}_{g_{i}}^{{ \; \:
 }^{n+1}},T_{i}^{n+1}) \; {\Delta \tilde{t}}_{g} { }^{n} \right] \; ,
 \end{array}
\end{equation} 
where $({\tilde{u}}_{i}^{n+1})_{\mboxs{ad}}$ and 
${\tilde{\rho}}_{g_{i}}^{{ \; \: }^{n+1}}$ are the specific internal
energy and the density which result from integrating the gas dynamics
equations without including the radiative terms, ${\mu}_{o}$ is the
mean molecular weight of a fully ionized primordial plasma, and 
$T_{\mboxs{vir}}$ is the virial temperature of the hosting halo,
\begin{equation}
T_{\mboxs{vir}} = 6 \; \times \; 10^{5} {\mu}_{o} \; {\left[ 
\frac{V_{c}}{100 \; \mbox{km s}^{-1}} \right]}^{2} \; \mbox{K} \; .
\end{equation}
The factor ${\cal L}$, 
\begin{equation} 
{\cal L} = 2.377 \times 10^{-23} \; (1 - Y_{\mboxs{P}})^{-2} \; h^{-1} 
{\left[ \frac{V_{c}}{100 \; \mbox{km s}^{-1}} \right]}^{2} \; , 
\end{equation} 
and the function $S({\rho}_{g},T)$, 
\begin{equation} 
S({\rho}_{g},T) \equiv \frac{\Gamma({\rho}_{g},T) -  
\Lambda({\rho}_{g},T)}{{n_{\mboxs{H}}}^{2}} \; , 
\end{equation} 
are both in units of erg s$^{-1}$ cm$^{3}$.

The main complication to find a solution to equation (29) is that  
the ionisation state of the plasma depends on the final temperature. 
We have designed an iteration scheme to find the temperature and  
the ionisation state at the end of the timestep. A first guess for  
the temperature is taken to be $T_{i}^{(1)} = T_{i}^{n}$. The rate  
equations are integrated, and the obtained fractional abundances   
are plugged in the right hand side of equation (29) to get a  
predicted temperature $T_{i}^{*}$. If the error function, 
\begin{equation} 
{\epsilon}_{i}^{(1)} = 1 - \frac{T_{i}^{*}}{T_{i}^{(1)}} \; ,
\end{equation} 
is smaller than 0.001 in absolute value, then $T_{i}^{n+1} =  
T_{i}^{*}$ and the integration is complete. If this is not the  
case a second guess for the temperature is taken to be, 
\begin{equation} 
T_{i}^{(2)} = \left[ 1 - 0.1 \; \mbox{sig}({\epsilon}_{i}^{(1)})  
\right] T_{i}^{(1)} \; ,
\end{equation} 
where $\mbox{sig}({\epsilon}_{i}^{(1)})$ is the signature of the  
error function. The rate equations are integrated and the whole  
process is repeated until the error function is smaller than 0.001,  
or changes sign. In the later case the root to equation (29) is  
bracketed with a 10\% error, and the Van Wijngaarden-Dekker-Brent  
root finder \cite{press} is used to find $T_{i}^{n+1}$ with  
a relative error smaller than 0.1\%.

\subsection{Inner boundary}

Once the gas has been able to cool efficiently it piles up
near the centre, and, since the calculations are performed in
a Lagrangian frame, the nodes of the grid get extremely concentrated.
This causes the code to require very small timesteps. Thoul \&
Weinberg \shortcite{twe1} got around this problem by freezing at a 
certain distance $r_{c}$ those zones for which the cooling
time has become much shorter than the dynamical time. However, they
found that the amount of cooled mass was sensitive to this central
boundary condition, and could vary by $\sim 10 \%$ depending on the
number of shells used. In addition, a recent numerical study of the 
stability of radiative shocks indicates that these zones play an
important role \cite{stb}. Therefore we keep the innermost zone
dynamically active, but we merge overlying zones with it whenever 
their temperature becomes smaller than a $T_{C} = 8000$~K.

%% file: sec3rad.tex
\subsection{Adiabatic collapse}

Bertschinger \shortcite{bert} found similarity
solutions for secondary infall and accretion onto a seed mass in an
Einstein de-Sitter universe. He studied not only collisonless and
collisional collapses, but also the collapse of non-self-gravitating
gas moving in the potential of a collisionless component. These 
solutions provide a good testing ground for the dynamics of our code 
in the absence of the cooling processes. In WZ's model, Bertschinger's
solutions correspond to $\epsilon = 1$ and no angular momentum (i.\
e.\ $e=1$). However, in order to avoid integration problems near the
centre we prefer to use slightly non-radial orbits for the dark matter
with an eccentricity $e=0.9$.

The self-similarity arises from the fact that there are no preferred
scales in the initial perturbation, and therefore, the whole evolution
of the system is determined only by the rate at which matter turns around. 
Following Bertschinger we nondimensionalize variables as follows:
\begin{equation}
 r = r_{t} \lambda
\end{equation}
\begin{equation}
 v(r,t) = \frac{r_{t}}{t} V( \lambda )
\end{equation}
\begin{equation}
 \rho(r,t) = {\rho}_{H} D( \lambda )
\end{equation}
\begin{equation}
 P(r,t) = {\rho}_{H} { \left( \frac{r_{t}}{t} \right) }^{2} P( \lambda
)
\end{equation}
\begin{equation}
 m(r,t) = \frac{4}{3} \pi {\rho}_{H} {r_{t}}^{3} M( \lambda )
\end{equation}
where ${\rho}_{H}$ and $r_t$ are the mean cosmic density and the
turnround radius at time t.

We first performed a calculation assuming a collisionless component 
and using $15000$ shells. This simulation settled down to self-similar
evolution which reproduced Bertschinger's solution almost perfectly 
except for small differences in the inner regions resulting from the 
fact that our shells have slightly nonradial orbits. The total energy
was conserved up to $0.3 \%$. We also calculated a collisionless
collapse with $\epsilon = 2/3$ and $e = 0.7$. Our numerical
solution agrees well with that given by White \& Zaritsky
\shortcite{wz}. The mass profile at redshift $z$ is linear in 
radius out to,
\begin{equation}
r_{\mboxs{vir}} \approx 120 \; {(1 + z)}^{-3/2} \; h^{-1} \; 
\left[ \frac{V_{\mboxs{c}}}{100 \; \mboxs{km s}^{-1}} \right] \; 
\mbox{kpc} \; ,
\end{equation}
and contains a mass,
\begin{equation}
M_{\mboxs{vir}} \approx 2.8 \times 10^{11} \; {(1 + z)}^{-3/2} \; 
h^{-1} \; {\left[ \frac{V_{\mboxs{c}}}{100 \; \mboxs{km s}^{-1}}
\right]}^{3} \; M_{\odot} \; .
\end{equation}
 
The self-similar solution for a collisional gas is obtained using a
grid with $500$ particles. We have simulated only the adiabatic case 
$\gamma = 5/3$, and find very good agreement with Bertschinger's 
similarity solution (see Table 1). The TAV scheme we have adopted 
proved to be very efficient in producing a sharp shock profile with 
practically no post-shock oscillations. We have also calculated a 
collisional collapse with both gas and a gravitationally dominant 
collisionless component (${\Omega}_{B} = 0.05$). In this case the
agreement with Bertschinger's solution is also good (see Table 2), 
but we observe some post-shock oscillations, especially in the
velocity profile. This is a spurious effect due to discrete jumps 
in the gravitational forces which occur whenever a dark matter shell 
crosses a gas zone. To reduce such effects we have to make sure that 
the shell masses are small enough compared to the gas zone masses. 
For equal masses this condition can be written as,
\begin{equation}
 \frac{N_{DM}}{N_{g}} \geq \frac{{\Omega}_{DM}}{{\Omega}_{g}} \; ,
\end{equation}
where $N_{DM}$, $N_{g}$ are the number of dark matter shells and gas 
zones respectively \cite{twe1}. The total energy was conserved up to
$0.2 \%$.

\subsection{Radiative processes}

In order to test our radiative processes subroutine we have calculated
the thermal evolution of the intergalactic medium (IGM) in the absence
of any perturbation. We assume, following  Miralda-Escud\'{e} \& Rees 
\shortcite{mer}, that the IGM was fully ionized at a redshift
$z_{in}$, and heated up to a temperature $T_{in}$. In their
calculations these authors took ${\Omega}_{\mboxs{B}} = 0.05$, and we 
have also studied the cases ${\Omega}_{b} = 0.01, \; 0.025 \;
\mbox{and} \; 0.1$. In the first case we find a good agreement with
the published results. For a lower ${\Omega}_{\mboxs{B}}$ the final
temperature of the IGM turns out to be lower, while for a higher 
baryonic content the final temperature is higher. This is the expected
behaviour since the recombination rates decrease with density, and 
consequently photoionization is not so effective at heating the gas.

We have also studied the thermal evolution of an overdense region, 
for which the density is given by the top-hat model until virialization, 
and is assumed constant thereafter. Once again the agreement with 
Miralda-Escud\'{e} \& Rees \shortcite{mer} is good. Now the final 
temperature of the cloud decreases as we increase the baryonic content
of the Universe. This is because at higher densities photoionization
is not effective anymore and the dominant process is line cooling,
which has an emissivity that increases with density.

%% file: sec4rad.tex
\subsection{Nonradiative collapse}

First of all we obtain the similarity solutions that arise from a
power-law initial perturbation with index $\epsilon = 2/3$ in the
absence of radiative cooling. If variables are scaled as in equations
(35)-(39) the partial differential equations which describe the gas
hydrodynamics transform into a set of ordinary differential equations,
\begin{equation}
(V - \lambda)D' + DV' + \frac{2}{\lambda}DV - 2D = 0 \; ,
\end{equation}
\begin{equation}
(V - \lambda)V' = - \frac{P'}{D} - \frac{2}{9} 
\frac{M}{{\lambda}^{2}} \; ,
\end{equation}
\begin{equation}
(V - \lambda) \left( \frac{P'}{P} - \gamma \frac{D'}{D} \right) = 2(1 -
\gamma) \; ,
\end{equation}
\begin{equation}
M' = 3 {\lambda}^{2} D \; ,
\end{equation}
where primes denote derivatives with respect to $\lambda$.

Far away from the origin the solution for cold accretion applies, 
in which case analytic expressions can be obtained (for details 
see Forcada-Mir\'{o} 1997),
\begin{equation}
 \lambda = sin^{2} \left( \frac{\theta}{2} \right) \; \frac{\pi }
{\theta - sin \theta} \; ,
\end{equation}
\begin{equation}
 V = \frac{\pi}{2} \; cot \left( \frac{\theta}{2} \right) \; ,
\end{equation}
\begin{equation}
 M = {\left( \frac{3 \pi}{4} \right)}^{2} \; \frac{\pi}{\theta - 
sin \theta} \; ,
\end{equation}
\begin{equation}
 D = \frac{1}{3} {\left( \frac{3 \pi}{4} \right)}^{2} \; 
\frac{sin^{- 2} \left( \frac{\theta}{2} \right) }{\lambda 
\left( \lambda - V \right) } \; ,
\end{equation}
where $\theta \in [0, \; 2 \pi]$. Turnaround corresponds to a 
value of the parameter ${\theta}_{\mboxs{t}} = \pi$.

At a given $\lambda = {\lambda}_{\mboxs{s}}$ the nondimensional 
strong shock jump conditions,
\begin{equation}
V_{2} - {\lambda}_{\mboxs{s}} = \left( \frac{\gamma - 1}{\gamma + 1}
\right) \left( V_{1} - {\lambda}_{\mboxs{s}} \right) \; ,
\end{equation}
\begin{equation}
D_{2} = \left( \frac{\gamma + 1}{\gamma - 1} \right) D_{1}
\end{equation}
\begin{equation}
P_{2} = \frac{2}{\gamma + 1} D_{1} {(V_{1} -
{\lambda}_{\mboxs{s}})}^{2} \; ,
\end{equation}
\begin{equation}
M_{2} = M_{1} \; ,
\end{equation}
provide the postshock values from which the ordinary differential 
equations (43)-(46) are solved using a Runge-Kutta integrator 
\cite{press}. The solution obtained for $\gamma = 5/3$ is 
plotted in Figure 2. There is only one position of the shock 
${\lambda}_{\mboxs{s}}^{o} \approx 0.29$ for which the boundary 
condition $V(0) = M(0) = 0$ is satisfied. In this case the solution
has an asymptotic behaviour near the origin, 
\begin{equation}
 D \sim {\lambda}^{-2}, \; \; P \sim {\lambda}^{-2}, \; \; M \sim 
\lambda \; .
\end{equation}
The density profile is due to the fact that shells settle at 
a constant fraction of their turnaround radius. The pressure
profile is a consequence of pressure forces balancing gravity in the
static regions. If the shock is at a smaller ${\lambda}_{\mboxs{s}}$ 
then gravity rapidly dominates over pressure gradients and the fluid 
approaches adiabatic free-fall as it flows inwards:
\begin{equation}
 V \sim {\lambda}^{-1/2}, \; \; D \sim {\lambda}^{-3/2}, \; \; P \sim
 D^{\gamma} \; .
\end{equation}
Each fluid particle crosses the origin at a finite time, and 
thus, there is a mass flux through the inner boundary; in the 
limit ${\lambda}_{\mboxs{s}} = 0$ the solution for cold accretion 
is recovered. On the other hand, if ${\lambda}_{\mboxs{s}} > 
{\lambda}_{\mboxs{s}}^{o}$ the mass goes to zero and the density 
diverges at a position $0 < {\lambda}_{o} < {\lambda}_{\mboxs{s}}$. 
Notice that only the solution with ${\lambda}_{\mboxs{s}} =
{\lambda}_{\mboxs{s}}^{o}$ is physical.

The introduction of a dominant collisionless component does not 
change the solution significantly from the purely gaseous case. 
We have performed the calculations for a baryonic content 
${\Omega}_{\mboxs{B}} = 0.05$ and eccentricities $e = 0.9, \; 
0.7, \; 0.5$. In the self-gravitating case the shock position is only
of the order of $1 \%$ smaller, and thus, the evolution in redshift of
$r_{\mboxs{s}}$ is well approximated for any baryonic content
${\Omega}_{\mboxs{B}} \in [0, \; 1]$ by the expression,
\begin{equation}
r_{\mboxs{s}} \approx 130 \; {(1 + z)}^{-3/2} \; h^{-1} \; 
\left[ \frac{V_{\mboxs{c}}}{100 \; {\mboxs{km s}}^{-1}} \right] \; 
\mbox{kpc} \; ,
\end{equation}
and the gas mass within,
\begin{equation}
M_{\mboxs{s}} \approx 3 \times 10^{11} {(1 + z)}^{-3/2} \; 
{\Omega}_{\mboxs{B}} h^{-1} \; {\left[ \frac{V_{\mboxs{c}}}
{100 \; {\mboxs{km s}}^{-1}} \right]}^{3} \; M_{\odot} \; .
\end{equation}
The mean overdensity enclosed by the shock is then ${\overline{D}}_{s}
\approx 120$, smaller than the usually assumed value of $200$ 
\cite{wf}. It is also interesting that the infalling gas hits the
shock with a velocity very close to the circular velocity of the 
inner halo.

\subsection{Radiative collapse}

The similarity solutions of the last section assume no radiative
cooling. However, if the virial temperature of the system is above 
the sharp cut-off of the cooling curve at approximately $10^4K$ the
post-shock gas will be able to radiate, and the evolution of the
system will be modified. According to equation (29) cooling should 
be important in halos with a circular velocity $V_{c} \geq 17$ 
km~s$^{-1}$. We have calculated the time evolution of systems with 
circular velocity 20, 40, 60, 80, 100, 140, 180, 220 and 300 
km~s$^{-1}$, and with gas fractions, ${\Omega}_{B} = 0.05$ and
$0.1$. Throughout this paper we adopt $H_{o} = 50$ km~s$^{-1}$ 
Mpc$^{-1}$.

Radiative cooling introduces a new scale length into the problem, 
the cooling radius, defined as the radius at which the cooling time 
equals the lifetime of the system. The self-similarity we had in the
nonradiative collapses is therefore broken. At a given epoch, the 
cooling radius and the shock radius are at different fractions of the
turnaround radius for each circular velocity. For a given circular 
velocity these fractions change with time (see 4.2.1 and 4.2.2).

The details of the behaviour of the radiative shock depend on 
whether the hosting halo has a circular velocity above or below 
$100$ km~$s^{-1}$. We therefore separate discussion of these two 
cases in the next two subsections. In the high circular velocity 
case the virial temperature is bigger than $3 \times 10^{5}$~K  and 
thermal bremsstrahlung dominates the cooling. Line emission by $H$ 
and $He^{+}$ provide the cooling processes in smaller halos.

\subsubsection{High circular velocity halos}

In Figure 3 we show the evolution of the radial structure  of a halo
like that of the Milky Way, i.e.~$V_{c} = 220$ km~s$^{-1}$. The
variables in this figure have been scaled as in equations (34-38); 
$\Theta$ is the temperature scaled to the virial temperature of the 
system (equation 29). For this example we have chosen $\Omega_b=0.05$.

At early times, the density profile inside the shock is well fitted by
a power-law exponent $-1.5$, the expected solution for a cooling flow
in an isothermal potential well \cite{bert2}. At later times, the 
profile has the exponent of the nonradiative model, $-2$, from the 
shock radius down to the cooling radius, and then at smaller radii 
it bends to an exponent of $-1.5$. The pressure profile is always
slightly steeper than the density profile, because residual departures
from hydrostatic equilibrium and the transition from the noncooling to
the cooling regime all cause the temperature to increase inwards. At
early times, the exponent for the pressure profile is roughly $-1.8$
over the regions we resolve, while at later times it is roughly $-2.3$
from the shock to the cooling radius, bending to $-1.8$ at smaller
radii. The mass profiles follows in the outer regions the adiabatic
collapse solution, and in the inner regions is dominated by the mass 
$M_{\mboxs{cool}}$ that has cooled below $8000$~K. At early times
$M_{\mboxs{cool}}$ is close to the mass of the black hole in the case
of cold accretion, and decreases as the system evolves in time. 

As gas passes through the cooling radius, radiative effects become
important; nevertheless, its temperature and density increase
continually until its cooling time becomes shorter than the local
dynamical time and catastrophic cooling takes place. The reason is
that as the gas radiates and flows inwards, compressional heating
dominates over radiative cooling. At the point where catastrophic 
cooling takes place, the material gets very cold and dense and piles 
up near the center. We cut our plots at this point because our code
cannot resolve the structure of the cold central region.

Another consequence of this inflow is that the gas is more centrally 
concentrated than the dark matter, which in principle could be a 
possible way of explaning the larger baryon fractions measured in 
rich galaxy clusters \cite{bk}. If we plot the baryon enhancement at 
a given time for different halos we see that the bigger the halo the 
closer we are to the adiabatic solution (Figure 4). On the other hand,
for each individual halo the global expansion of the Universe reduces 
all densities and so makes radiative cooling progressively less
efficient with time. As a result the gas enhancement gets weaker 
as the system evolves. This explains why the profile of enclosed gas 
mass in Figure 3 gets steeper with time. However, White \itl{et al.} 
\shortcite{wn} proved that even assuming the most extreme case of cold
accretion one cannot account for the observed baryon fractions
in clusters of galaxies. Cooling processes alone cannot therefore 
enhance the cluster baryon fraction in an Einstein-de Sitter
Universe to the values required for consistency with 
$\Omega_B$ as inferred from Big Bang nucleosynthesis.

In Figure 5 we present the evolution of the shock radius and 
the cooling radius for circular velocities 140, 180, 220 and 
300 km~s$^{-1}$, and in Figure 6 the evolution of the mass within 
these radii. In these simulations the central resolution is such 
that the shock forms at a redshift $z \sim 50$. At early times all 
the gas that goes through the shock is able to radiate away most of 
its internal energy. As a result, the shock wave and the cooling wave 
move together, their position increasing approximately as $\sim
t^{\alpha}$, with $\alpha$ decreasing slightly with circular
velocity. At later times the cooling wave separates from the shock
wave, because the post-shock density is no longer high enough for 
rapid cooling, and the shock position and the mass it contains grow in time 
in proportion to $t^{1/2}$, as in Bertschinger's (1986) similarity
solution. The shock wave asymptotically approaches
the nonradiative regime, its position and mass growing linearly in
time. Figure 6 also shows the total mass that has cooled below 8000~K,
and so is included in the central zone.

\subsubsection{Low circular velocity halos}

The behaviour of the radiative shocks in systems with $V_{c} < 
100$ km~s$^{-1}$ differs significantly from that in higher circular
velocity halos. In order to clarify the situation in this case we 
have followed the evolution over longer timescales, corresponding 
to epochs beyond $z=0$. In order to do this it was necessary to 
reduce the central resolution slightly; the shock then forms at a 
redshift $z \sim 25$. 

The most noticeble feature of these simulations is that the shock
front is unstable to large amplitude oscillations when $20$ km~s$^{-1}
< V_{c} < 80$ km~$s^{-1}$ (Figure 7). The onset of these oscillations 
takes place earlier with decreasing circular velocity. Their amplitude
and period grow as the system gets older, and, in the cases with 40
and 60 km~$s^{-1}$, they die out at later times. The oscillations can 
also be seen in plots of the mass within the cooling and shock radii 
(Figure 8). It is interesting that during most of the oscillatory 
evolution no gas cools below $8000$~K; only at the very end of the 
cycle does strong cooling allow a burst of mass accretion onto the 
central object.

Radiative shocks have been found to be unstable to oscillations in 
other astrophysical contexts. Langer, Chanmugam \& Shaviv \shortcite{lcs} 
simulated time-dependent accretion onto white dwarfs and found that
the cooling layer suffers oscillatory instabilities if the rate of 
accretion is above a certain critical value. Improved hydrodynamical 
simulations showed that these radiative shocks are unstable to
oscillations if the cooling rate $\Lambda \sim {\rho}^{2}T^{c}$ 
has a dependence on temperature such as $\alpha < 0.6$, in quite 
good agreement with linear stability analysis \cite{iwd}.

These studies considered equilibrium ion abundances, and in our 
calculation we use time-dependent cooling. Therefore, our cooling 
function is dependent on the history of the plasma. Gaetz, Edgar \& 
Chevalier \shortcite{gec} addressed the question of stability of 
radiative shocks with time-dependent cooling and concluded that, 
although a precise stability limit cannot be established, simple 
power-law approximations of the cooling curve give a rough idea 
of the driving physics.

With a power-law approximation the cooling function, the cooling 
time of a gas element goes as,
\begin{equation}
t_{\mboxs{cool}} \sim {{\rho}_{\mboxs{g}}}^{-1} T^{1 - c} \; .
\end{equation}
The post-shock temperature $T_{\mboxs{s}}$ of the gas is related 
to the pre-shock infall velocity $v_1$ and the shock velocity 
$v_{\mboxs{s}}$ through the strong shock jump conditions:
\begin{equation}
 T_{s} \sim \; {\left( v_{1}- v_{\mboxs{s}} \right)}^{2} \; .
\end{equation}
If the shock moves outwards the gas is heated to a higher temperature,
while the post-shock density drops (Figure 9). This leads to an
increased cooling time, particularly for small $c$, and so to
increased pressure in the region behind the shock. This excess
pressure drives the shock further outwards in a very short
timescale. In this expansion phase the density, temperature and
pressure profiles are rather featureless, with slopes close to the
cooling regime discussed in Section 4.2.1 (Figure 10).

Eventually, the density of the downstream gas increases to the point 
where cooling again becomes substantial; pressure support for the
shock then weakens and the shock slows down. The gas is now heated 
to a smaller temperature and can cool more efficiently (Figure 9). 
As a result of this cooling instability a density peak forms behind 
the shock (Figure 10). The rapid cooling of the gas in the density 
peak produces a further loss of pressure support and the shock falls 
in with the flow. It is during this final collapse that cold gas is 
accreted onto the centre during a relatively short ``burst''. A 
pressure build-up near the centre eventually causes the infalling 
shock to strengthen again and to reverse its motion. As a result 
of the cosmic expansion, it takes longer for the post-shock gas 
to cool in each new cycle, causing the period and amplitude of 
the oscillations to increase.

In our models we see such oscillatory behaviour when $V_{\mboxs{c}} 
\leq 80$ km~s$^{-1}$. This suggests that the critical slope of the 
cooling function for instability is $c \sim 0$. For the primordial 
element abundances which we assume, $c$ is reduced below the value
$c = 0.5$ expected for bremsstrahlung as helium line cooling
enhances the radiation at lower temperatures. Including heavier
elements in our cooling function would cause the transition to 
oscillatory behaviour to move to higher $V_{\mboxs{c}}$.

Another interesting result is that in these low circular 
velocity halos the gas component is more concentrated for a given
fraction of the turnaround radius than in bigger systems (Figure 4). 
However, observations of dwarf galaxies seem to indicate that they
are dark-matter dominated \cite{kor}. Therefore, one needs some
physical mechanism to dilute the baryon component in the smaller 
systems if one wants to reconcile this theory of galaxy formation 
with observations. Dekel \& Silk \shortcite{ds} roughly estimated 
that a starburst could cause loss of most of the gas in a halo with circular 
velocity below a critical value of the order of $100$km~s$^{-1}$. Their 
proposed mechanism is supernova-driven winds, since gas absorbs 
radiation emitted by young stars very inefficiently. It is quite 
uncertain how much mass is lost by this mechanism which must
be effective well before most of the gas turns into stars.

%% file: sec5rad.tex
We desire to present physical explanations for the evolution in time 
of the shock and cooling wave. In order to accomplish this, we have
used as a starting point the ideas set out in White \& Frenk 
\shortcite{wf}. They pointed out that due to the general expansion 
of the Universe the shock will go through two very distinct epochs:
\begin{enumerate}
 \item Infall-dominated phase: Initially the density is high enough 
       that all the gas that gets shock-heated is able to cool (i.e.\ 
       $t_{\mboxs{cool}} < t$, where t is the lifetime of the system). 
       At this time the supply of cold gas is limited by the infall 
       rate rather than by cooling.
 \item Cooling-dominated phase: After a certain time the density
       becomes too low and the post-shock gas cannot cool efficiently 
       anymore (i.e.\ $t_{\mboxs{cool}} > t$). At this stage the
       amount of cold gas is regulated by radiative losses.
\end{enumerate}
These two epochs have been clearly seen in the numerical simulations
presented in Section 4, and we will study them now in more detail.

\subsection{Infall-dominated regime}

During the infall-dominated phase the gas is able to cool so 
efficiently that it rapidly loses pressure support. Consequently it 
flows towards the centre, most of the cooling layer being covered 
while the matter is still travelling at the postshock velocity. 
The thickness of the cooling layer is then roughly,
\begin{equation}
{\Delta r}_{\mboxs{cool}} \sim v_{\mboxs{s}} \; t_{\mboxs{cool}} \; .
\end{equation}
As the gas cools it is compressed to high density. The whole process
can be regarded as an isothermal shock, and thus the compression
factor is,
\begin{equation}
{\rho}_{2} \approx {\cal M}_{s}^{2} \; {\rho}_{1}
\end{equation}
where ${\cal M}_{s}$ is the Mach number of the shock,
\begin{equation}
{\cal M}_{s} \approx {\left( \frac{\mu \; m_{\mboxs{p}} \; 
v_{\mboxs{s}}^{2}}{\gamma \; k_{\mboxs{B}} \; T_{\mboxs{c}}} 
\right)}^{1/2} \gg 1 \; .
\end{equation}
and $T_{\mboxs{c}}$ is the gas temperature, which is of the order of 
the cut-off temperature in the cooling curve (i.e.\ $T_{\mboxs{c}} 
\sim 10^{4}$~K). The thickness of the cold layer can be roughly
estimated from mass conservation,
\begin{equation}
{\Delta r}_{\mboxs{cold}} \approx {\cal M}_{\mboxs{s}}^{-2/3} \;
r_{\mboxs{s}} \; .
\end{equation}
Therefore, the shock will be at a position,
\begin{equation}
r_{\mboxs{s}} \approx {\Delta r}_{\mboxs{cool}} \gg 
{\Delta r}_{\mboxs{cold}} \; ,
\end{equation}
which implies that the shock moves in the quasi-equilibrium,
\begin{equation}
t_{\mboxs{cool}} \sim  t_{\mboxs{ff}} \; ,
\end{equation}
where $t_{\mboxs{ff}}$ is the freefall timescale, 
\begin{equation}
 t_{\mboxs{ff}} = \frac{1}{\sqrt{24 \pi G \overline{\rho}}} \; 
\sim \; {\left( \frac{M_{\mboxs{s}}}{{\lambda}_{\mboxs{s}}^{3}}
 \right)}^{-1/2} \; {(1 + z)}^{-3/2} \; ,
\end{equation}
and $\overline{\rho}$ is the mean density of total mass within a
given radius. If the cooling time is bigger than the free-fall time 
adiabatic heating will increase the temperature of the gas thereby reducing 
the cooling time. In the opposite situation the temperature
rapidly decreases, and thus, the cooling time increases.

At very early times Compton scattering provides the dominant source of
energy loss \cite{si}. In this case, the cooling time depends only on 
the redshift,
\begin{equation}
t_{\mboxs{Comp}} \approx 2.4 \times 10^{12} \; {(1 + z)}^{-4} \;
\mbox{yr} \; .
\end{equation}
According to equation (66), the evolution of the shock is
qualitatively given by,
\begin{equation}
\frac{M_{\mboxs{s}}}{{\lambda}_{\mboxs{s}}^{3}} \sim {(1 + z)}^{5} 
\; .
\end{equation}
The gas flows inwards until it becomes self-gravitating, 
which implies that $M_{\mboxs{s}}$ is roughly a constant,
\begin{equation}
M_{\mboxs{s}} \approx {\Omega}_{\mboxs{B}} M_{o} \; ,
\end{equation}
where $M_{o}$ is the nondimensional mass ***what does this mean?***
of the black hole in the 
cold collapse solution. Therefore, the position of the shock evolves 
in time as,
\begin{equation}
r_{\mboxs{s}} \sim {\Omega}_{\mboxs{B}}^{1/3} t^{19/9} \; .
\end{equation}

The rate of energy loss due to Compton scattering decreases with a
high power of the redshift, and eventually two-body processes will
dominate the radiative cooling of the system. In this case the
cooling time depends on the nondimensional hydrodynamic variables
as, 
\begin{equation}
t_{\mboxs{cool}} \; \sim \;  {D_{1}}^{-1} \; {V_{1}}^{2(1 - c)} 
\; {(1 + z)}^{-3} \; ,
\end{equation}
and the qualitative evolution of the radiative shock is given by,
\begin{equation}
\Psi(\lambda ; c , {\Omega}_{\mboxs{B}}) \equiv 
D_{1} {V_{1}}^{-2(1 - c)} {\left( 
\frac{M_{\mboxs{s}}}{{\lambda}_{\mboxs{s}}^{3}} 
\right)}^{-1/2} \sim {(1 + z)}^{-3/2} \; ,
\end{equation}
where the function $\Psi(\lambda ; c , {\Omega}_{\mboxs{B}})$ 
depends on two parameters: the slope of the cooling function $c$, 
and the baryonic content of the universe ${\Omega}_{\mboxs{B}}$. 
For a self-gravitating gas one has the asymptotic behaviour,
\begin{equation}
\Psi(\lambda \ll 1 ; c , {\Omega}_{\mboxs{B}}) \approx 
\Psi_{o}({\Omega}_{\mboxs{B}}) \; {\lambda}^{1 - c} \; ,
\end{equation}
where $\Psi_{o}({\Omega}_{\mboxs{B}})$ increases as the baryonic
fraction decreases, and diverges for ${\Omega}_{\mboxs{B}} \rightarrow
0$. Therefore, the physical position of the radiative shock has a 
power-law behaviour in time,
\begin{equation}
r_{\mboxs{s}}(t ; c , {\Omega}_{\mboxs{B}}) \approx
r_{o}({\Omega}_{\mboxs{B}}) \; t^{\frac{2 - c}{1 - c}} \; ,
\end{equation}
where $r_{o}({\Omega}_{\mboxs{B}})$ decreases with the baryonic
fraction, and $r_{o}({\Omega}_{\mboxs{B}}) \rightarrow 0$ as 
${\Omega}_{\mboxs{B}} \rightarrow 0$.

The radius of the shock wave in the infall-dominated regime grows
faster than linearly in time; as a result the nondimensional 
shock position ${\lambda}_{\mboxs{s}}$ increases as the system 
evolves. The slope of the function $\Psi(\lambda ; c ,
{\Omega}_{\mboxs{B}})$ increases with $\lambda$, and thus, the
power-law behaviour of $r_{\mboxs{s}}$ decreases approaching a 
linear growth in time (Figure 11). Eventually, the postshock cooling 
time becomes bigger than the lifetime of the system, and the shock 
enters the cooling-dominated regime. An upper limit to the position 
at which this transition happens is provided by the nonradiative 
shock (i.e.\ ${\lambda}_{\mboxs{s}} \approx 0.29$). At this point,
$r_{\mboxs{s}}$ has roughly a power-law behaviour,
\begin{equation}
\alpha \approx \frac{1.5 - c}{1 - c} \; .
\end{equation}
Therefore, if one fits a single power-law to the time evolution of 
the radiative shock one should obtain a value in between these two
limits.

One has to keep in mind that the softened gravity prevents the
collapse beyond the resolution limit of the code, and the shock 
forms at a position close to the softening parameter $a$. For 
$r < a$ a cold dense core forms, where the density is constant in
radius and the material is at rest. At this stage the shock strongly
violates the quasi-equilibrium condition, with $t_{\mboxs{cool}} \ll
t_{\mboxs{ff}}$. In the numerical simulations which do not present an
oscillatory instability we have fitted an analytical expression of the
form $r_{\mboxs{s}} = a + b t^{\alpha}$ to the evolution in time of
the shock radius (Figure 5 and Figure 7). The fitted exponent is in
all cases between the predicted limits, and for the high velocity
halos is fairly constant, with a value close to $\alpha = 2$. This is
the exponent expected when Compton cooling is still a significant source
of cooling (Figure 12).

\subsection{Cooling-dominated regime}

The transition to the cooling-dominated regime happens when the
postshock cooling time equals the lifetime of the system. According to
equations (66) and (69) the mean total overdensity has a well defined
value at this time. This is in agreement with the fact that in the
simulations the nondimensional position at which the shock transits is
the same for all halos with a very small scatter (Table 3). Also the
postshock temperature is very close to the virial temperature of the
system. 

The nondimensional position of the shock at the time of transition is
very close to the case $\gamma = 4/3$. In fact, radiative cooling
weakens the pressure support of the gas against gravity, making the
equations of state less stiff; in other words, radiative cooling leads
to an effective ratio of specific heats ${\gamma}_{\mboxs{eff}} <
5/3$, and, if the gas cools strongly, even ${\gamma}_{\mboxs{eff}} <
4/3$. The numerical simulations suggest that one can think of the
evolution in time of a collapsing halo as ``passing'' through a series
of adiabats. At early times, when radiative cooling is very strong,
one has ${\gamma}_{\mboxs{eff}} < 4/3$. This exponent increases with
time towards $\gamma
= 5/3$. The value $\gamma = 4/3$ delimits the boundary between the
infall-dominated and the cooling-dominated regime.

With the appropriate choice of temperature and gas density one can
obtain a fairly good estimate of the redshift $z_{\mboxs{cf}}$ at
which the cooling-dominated phase starts. If one takes a temperature
equal to the virial temperature of the halo, and a density equal to
the mean gas overdensity inside the nonradiative shock the predictions
compare well with the results from the simulations (Figure 13). Now,
for a given halo, the physical size at the time the transition takes 
place is simply,
\begin{equation}
R_{\mboxs{cf}} \approx 10 \; {\left[ \frac{1 + z_{\mboxs{cf}}}
{1 + z_{\mboxs{cf}}^{*}} \right]}^{-3/2} \; h^{-1} \; 
\left[ \frac{V_{\mboxs{c}}}{100 \; \mbox{km s}^{-1}} \right] \; 
\mbox{kpc} \; ,
\end{equation}
where $z_{\mboxs{cf}}^{*} = 3$ is a characteristic redshift, and the
total mass of cold gas enclosed in solar masses is,
\begin{equation}
M_{\mboxs{cf}} \approx 3.8 \times 10^{10} {\left[ \frac{1 + 
z_{\mboxs{cf}}}{1 + z_{\mboxs{cf}}^{*}} \right]}^{-3/2}  
{\Omega}_{\mboxs{B}} h^{-1} {\left[ \frac{V_{\mboxs{c}}}
{100 \; \mbox{km s}^{-1}} \right]}^{3} \; .
\end{equation}

The simulations show that as soon as the shock enters the 
cooling-dominated regime a hot halo starts building with an 
isothermal density profile. The temperature has a slight 
increase inwards (Figure 3), but we will assume it is a constant and 
equal to the virial temperature of the hosting halo. At any given time 
there is a radius $r_{\mboxs{c}} \le r_{\mboxs{s}}$ at which the gas 
density is high enough so the cooling time equals the life time of the 
system,
\begin{equation}
t \approx t_{\mboxs{cool}}(r_{\mboxs{s}}) \; 
\frac{{\rho}_{\mboxs{g}}(r_{\mboxs{s}})}
{{\rho}_{\mboxs{g}}(r_{\mboxs{c}})} \; ,
\end{equation}
and since the halo is isothermal,
\begin{equation}
r_{\mboxs{c}}(t) \approx r_{\mboxs{s}}(t) \; {\left[
\frac{t}{t_{\mboxs{cool}}(r_{\mboxs{s}})} \right]}^{1/2} \; .
\end{equation}
Once the shock has entered the cooling-dominated regime, it gradually
approaches the nonradiative behaviour and grows linearly in time. At 
this stage the postshock cooling time grows proportional to $t^{2}$, 
and then,
\begin{equation}
r_{\mboxs{c}}(t) \approx r_{\mboxs{s}}(t_{\mboxs{o}}) \; 
{\left[ \frac{t}{t_{\mboxs{cool}}^{*}} \right]}^{1/2} \; ,
\end{equation}
where  $t_{\mboxs{o}}$ is a reference epoch at which the shock 
presents the nonradiative behaviour, and $t_{\mboxs{cool}}^{*}$ 
is the post-shock cooling time at the reference epoch. 

The cooling time $t_{\mboxs{cool}}^{*}$ is related to the estimated 
time at which the shock enters the cooling dominated regime 
$t_{\mboxs{cf}}$ as,
\begin{equation}
\frac{t_{\mboxs{cool}}^{*}}{t_{\mboxs{cf}}} = 
\frac{{\overline{D}}_{\mboxs{s}}}{D_{\mboxs{s}}} \;
{\left[ \frac{t_{\mboxs{o}}}{t_{\mboxs{cf}}}\right]}^{2} \; ,
\end{equation}
where ${\overline{D}}_{\mboxs{s}}$ is the mean gas overdensity inside
the nonradiative shock and $D_{\mboxs{s}} \approx 27$ is the self-similar 
postshock density. Combining equations (81) and (82) we have obtained
the following analytic approximation for the evolution in time of the 
position of the cooling wave, 
\begin{equation}
r_{\mboxs{c}}(t) \approx {\left[
\frac{D_{\mboxs{s}}}{{\overline{D}}_{\mboxs{s}}} \right]}^{1/2}  
r_{\mboxs{s}}(t_{\mboxs{cf}}) {\left[ \frac{t}{t_{\mboxs{cf}}} 
\right]}^{1/2} \approx \frac{1}{2} r_{\mboxs{s}}(t_{\mboxs{cf}}) 
{\left[ \frac{t}{t_{\mboxs{cf}}} \right]}^{1/2} \; .
\end{equation}

In Figure 5 and Figure 7 one can see that the analytical expression
given in equation (83) is a rather good approximation to the position 
of the cooling wave once the shock wave presents the nonradiative 
behaviour. Until this happens the postshock density decreases in time 
faster than in the self-similar nonradiative evolution, so the cooling
radius will be at a smaller fraction of the shock radius than
predicted by a power-law scaling. In fact, in Figure 5 one can see 
that in the early stages of the cooling flow the radius
$r_{\mboxs{c}}$ decreases with time. On the other hand, the cooling 
radius cannot move inwards faster than the flow, so it might happen 
that the cooling wave is stationary at the early stages, with no 
mass accretion. The numerical code only gives the position of the 
cooling radius whenever a cell can first cool efficiently, and thus, 
a stationary cooling wave will produce a ``gap'' in the data. This 
``gap'' is clearly seen for $V_{\mboxs{c}} = 300$ km~s$^{-1}$ in both 
the radius (Figure 5) and the mass evolution (Figure 6), and barely 
for $V_{\mboxs{c}} = 220$~km~s$^{-1}$. This effect is stronger with 
increasing circular velocity because the power-law $\alpha$ is 
bigger (Figure 12).

In the self-similar solutions of Bertschinger (1989) the mass accretion 
rate at the cooling radius is approximately,
\begin{equation}
{\dot{M}}_{\mboxs{c}} \approx 4 \; \pi \; {r_{\mboxs{c}}}^{2} \; 
{\rho}_{\mboxs{g}}(r_{\mboxs{c}}) \; {\dot{r}}_{\mboxs{c}} \; .
\end{equation}
so we model the rate at which cold gas accumulates by this simple 
inflow equation. Since the halo is isothermal down to the cooling radius,
\begin{equation}
{r_{\mboxs{c}}}^{2} \; {\rho}_{\mboxs{g}}(r_{\mboxs{c}}) = 
{r_{\mboxs{s}}}^{2} \; {\rho}_{\mboxs{g}}(r_{\mboxs{s}}) \; ,
\end{equation}
and at later times, when the system is well into the cooling-dominated 
regime, the total mass accreted by the shock is,
\begin{equation}
M_{\mboxs{s}} \approx 4 \; \pi \; {r_{\mboxs{s}}}^{3} \; 
{\rho}_{\mboxs{g}}(r_{\mboxs{s}}) \; .
\end{equation}
Now we can write equation (84) as,
\begin{equation}
{\dot{M}}_{\mboxs{c}} \approx \frac{M_{\mboxs{s}}}{r_{\mboxs{s}}} \; 
{\dot{r}}_{\mboxs{c}} \; .
\end{equation}
According to equation (87), and since the shock evolves like in the 
nonradiative case, the mass inside the cooling radius grows in time 
proportional to $t^{1/2}$. In Figure 6 and Figure 8 one can see
that a simple power-law scaling,
\begin{equation}
M_{\mboxs{c}}(t) \approx M_{\mboxs{s}}(t_{\mboxs{cf}}) \; {\left[
\frac{t}{t_{\mboxs{cf}}} \right]}^{1/2} \; ,
\end{equation}
is a good approximation for the mass within the cooling radius.

%% file: sec6rad.tex
1-D simulations like the ones we have done are a powerful tool to
enhance our understanding of the physics involved in the process of
galaxy formation. In particular, we have concentrated on the shocks 
that appear in the collapse of a protogalaxy, and they have been
proved to present the features already known from previous studies of 
strongly radiative shocks in other contexts. The shocks are subject to
an oscillatory instability whenever line cooling dominates the
radiative losses behind the shock front. Due to the expanding
background in which the system is embedded, the period and the
amplitude of the oscillations increase with time.

We believe that we now fully understand the effects of radiative cooling 
in the shocks that appear in spherical models for galaxy formation. Despite
the rather complicated physical behaviour of the shocks, we have
been able to build a simple model which explains very well the
kinematics both of the shock wave and of the cooling wave in all cases
which are not unstable. At early times the shock front evolves in such
a way that the cooling time is of the order of the free-fall time. In
the bigger halos Compton cooling dominates, and the radius of the
shock grows proportional to $t^{2}$. In smaller halos the slope is a
function of the slope of the cooling function at the virial
temperature of the system. The transition to the cooling-dominated
regime happens at the same nondimensional position in all halos. This
is close to the position of a nonradiative shock with a ratio of
specific heats $\gamma = 4/3$. The time at which each system becomes
nonradiative is well estimated assuming a temperature equal to the
virial temperature of the system, and an overdensity equal to the mean
overdensity inside the nonradiative shock.

Whenever oscillations appear, our model provides only a rough guide to
the behaviour. However, one should keep in mind that there are
important processes that we have not taken into 
account, and that affect systems with circular
velocities precisely in the range where the shock wave oscillates. The cold 
dense gas that accumulates in the central regions will form stars, 
which will provide a flux of ionizing radiation. In addition we should
consider a metagalactic background of UV radiation due to high
redshift quasars. The effect of such  
radiation is to maintain a higher ionization fraction which
reduces the line cooling. We may expect the oscillatory behaviour
to be absent in this case, and lower circular velocity dwarf galaxies
be prevented from forming (Efstathiou 1992). We will consider such
effects further in a future paper. They have convinced us that 
further analysis of the oscillatory solutions is unlikely to be
worthwhile.

%% file: fig01rad.tex
\clearpage 

\begin{figure*}
   \cidfig{17cm}{20}{160}{570}{700}{\DIRFIGS time.ps}
\vspace{6pt}{
{\bf Figure 1:} (a) Coulomb equilibration time (straight solid line),
cooling time (solid), and collisional ionisation times for HI
(dotted),  HeI (short-dashed) and HeII (long-dashed) for a primordial 
plasma with  ${\Omega}_{\mboxs{B}} = 0.05$, $h = 0.5$ and residual 
ionised fraction taken from Peebles (1993). (b) Cooling time (solid) 
and recombination times for HII (dotted), HeII (short-dashed) and 
HeIII (long-dashed) for a primordial plasma with ${\Omega}_{\mboxs{B}}
= 0.05$, $h = 0.5$ and equilibrium ionised fraction.
}
\end{figure*}

%% file: fig02rad.tex
\begin{figure*}
   \cidfig{17cm}{20}{150}{580}{700}{\DIRFIGS sim1.ps}
\vspace{6pt}{
{\bf Figure 2:} Self-similar velocity (top-left), density (top-right),
pressure (bottom-left), mass (bottom-right) profiles for shocked
accretion in the case $\epsilon = 2/3$ and $\gamma = 5/3$. Solid line:
solution satisfying the inner boundary V(0) = M(0) =0. Dashed lines:
solutions with ${\lambda}_{\mboxs{s}} < {\lambda}_{\mboxs{s}}^{o}$ or 
${\lambda}_{\mboxs{s}} > {\lambda}_{\mboxs{s}}^{o}$, where 
${\lambda}_{\mboxs{s}}^{o}$ is the position of the shock for the
solution plotted in solid line. The expected asymptotic slopes are
indicated.
}
\end{figure*}

%% file: fig03rad.tex
\begin{figure*}
   \cidfig{17cm}{20}{20}{580}{750}{\DIRFIGS profc.ps}
\vspace{6pt}{
{\bf Figure 3:} Radial profiles at different redshifts for a halo
with $V_{\mboxs{c}} = 220$ km~s$^{-1}$. Top: Density. The two slopes
plotted are -2 and -1.5. Middle: Temperature. The slope plotted is
-0.3. Bottom: Mass. The top horizontal dashed line is the mass of the 
black hole in the case of cold accretion and the bottom one is the total
mass that has cooled below 8000 K. The solid line is the self-similar
mass profile in the case of adiabatic collapse. In all cases the left 
vertical  dashed line is the position of the cooling wave and the 
right one is the position of the shock wave.
}
\end{figure*}

%% file: fig04rad.tex
\begin{figure*}
   \cidfig{17cm}{20}{140}{570}{700}{\DIRFIGS enh.ps}
\vspace{6pt}{
{\bf Figure 4:} Mass of baryons to total mass ratio versus fraction
of the turnaround radius at $z = 0$. In short-dashed line, from left
to right one has halos with circular velocity $300$, $140$, $80$, $40$
and $20$ km~$s^{-1}$. In solid line one has the adiabatic collapse
(bottom) and collapse onto a black hole (top). In all cases
${\Omega}_{\mboxs{B}} = 0.1$.
}
\end{figure*}

%% file: fig05rad.tex
\begin{figure*}
   \cidfig{17cm}{10}{140}{580}{700}{\DIRFIGS shock2.ps}
\vspace{6pt}{
{\bf Figure 5:} Evolution in time of the shock wave position (solid
line) and the cooling wave position (crosses) as given by the
numerical simulations for halos with $V_{\mboxs{c}} > 100$
km~s$^{-1}$ and ${\Omega}_{\mboxs{B}} = 0.05$. Long-dashed lines 
correspond to the adiabatic evolution of the shock. Dotted-lines
correspond to a growth in time proportional to $t^{\alpha}$, and an
off-set equal to the softening parameter. The value of the exponent
$\alpha$ is given in each panel. Short-dashed lines correspond to the
analytical expression given in equation (6.31). 
}
\end{figure*}

%% file: fig06rad.tex
\begin{figure*}
   \cidfig{17cm}{10}{140}{580}{700}{\DIRFIGS mass2.ps}
\vspace{6pt}{
{\bf Figure 6:} Evolution in time of the gas mass within the shock
wave (solid line), the cooling wave (crosses), and the mass that has
cooled below $8000$ K (dotted line) as given by the numerical
simulations for halos with $V_{\mboxs{c}} > 100$ km~s$^{-1}$ and 
${\Omega}_{\mboxs{B}} = 0.05$. Long-dashed lines correspond to the 
adiabatic evolution of the shock. Short-dashed lines indicate a growth
in time proportional to $t^{-1/2}$. 
}
\end{figure*}

%% file: fig07rad.tex
\begin{figure*}
   \cidfig{17cm}{10}{140}{580}{700}{\DIRFIGS shock1.ps}
\vspace{6pt}{
{\bf Figure 7:} Evolution in time of the shock wave position (solid
line) and the cooling wave position (crosses) as given by the
numerical simulations for halos with $V_{\mboxs{c}} < 100$
km~s$^{-1}$ and ${\Omega}_{\mboxs{B}} = 0.05$. Long-dashed lines 
correspond to the adiabatic evolution of the shock. In the bottom
right panel the short-dashed line corresponds to the analytical
expression given in equation (6.31). The dotted-lines indicate a 
power-law $t^{1.5}$, with an off-set equal to the softening 
parameter.
}
\end{figure*}

%% file: fig08rad.tex
\begin{figure*}
   \cidfig{17cm}{10}{140}{580}{700}{\DIRFIGS mass1.ps}
\vspace{6pt}{
{\bf Figure 8:} Evolution in time of the gas mass within the shock
wave (solid line), the cooling wave (crosses), and the mass that has
cooled below $8000$ K (dotted line) as given by the numerical
simulations for halos with $V_{\mboxs{c}} < 100$ km~s$^{-1}$ and 
${\Omega}_{\mboxs{B}} = 0.05$. Long-dashed lines correspond to the 
adiabatic evolution of the shock. In the bottom right panel the
short-dashed line corresponds to a growth in time proportional to
$t^{1/2}$.
}
\end{figure*}

%% file: fig09rad.tex
\begin{figure*}
   \cidfig{17cm}{20}{140}{580}{700}{\DIRFIGS cycle.ps}
\vspace{6pt}{
{\bf Figure 9:} Evolution in time of the post-shock density (top)
and the post-shock temperature (bottom) for a halo with $V_{\mboxs{c}}
= 40$ km~s$^{-1}$ and ${\Omega}_{\mboxs{B}} = 0.05$. Short-dashed
lines indicate the times at which radial profiles are plotted in
Figure 6.9.
}
\end{figure*}

%% file: fig10rad.tex
\begin{figure*}
   \cidfig{17cm}{20}{140}{580}{700}{\DIRFIGS snap.ps}
\vspace{6pt}{
{\bf Figure 10:} Radial profiles at different redshifts for a halo
with $V_{\mboxs{c}} = 40$ km~s$^{-1}$ and ${\Omega}_{\mboxs{B}} =
0.05$. Top: Density. Middle: Temperature. Bottom: Pressure. The left
panel is a snapshot taken just after the shock reached maximum
expansion. The right panel was taken during the expansion phase. The
short-dashed lines indicate the power-law behaviour $-1.5$, $-0.3$,
$-1.8$ for the density, temperature and pressure respectively.
}
\end{figure*}

%% file: fig11rad.tex
\begin{figure*}
   \cidfig{17cm}{10}{130}{580}{700}{\DIRFIGS black2.ps}
\vspace{6pt}{
{\bf Figure 11:} Variation of the slope of $r_{\mboxs{s}}(t)$ with 
$\lambda$ for different slopes of the cooling function $c$ in the case
${\Omega}_{\mboxs{B}} = 1$. The long-dashed lines indicate the
position of the nonradiative shock and the turnaround radius.
}
\end{figure*}

%% file: fig12rad.tex
\begin{figure*}
   \cidfig{17cm}{5}{5}{550}{550}{\DIRFIGS slope.ps}
\vspace{6pt}{
{\bf Figure 12:} Solid lines: Predicted limits for the 
power-law evolution in time of the position of a radiative shock 
$r_{\mboxs{s}} \sim t^{\alpha}$ in the case two-body process dominate
the radiative cooling. The lower limit is the analytical approximation
to the numerical limit (dotted line). The short-dashed line is the
expected exponent if Compton cooling is the dominant cooling. Filled 
squares: Fitted exponent to simulations with ${\Omega}_{\mboxs{B}} = 
0.05$. Empty squares: Fitted exponent to simulations with 
${\Omega}_{\mboxs{B}} = 0.1$.
}
\end{figure*}

%% file: fig13rad.tex
\begin{figure*}
   \cidfig{17cm}{10}{140}{580}{710}{\DIRFIGS coolpre.ps}
\vspace{6pt}{
{\bf Figure 13:} Predicted redshift for the onset of a cooling flow
at the central regions of a collapsing halo in a flat unvierse with a
baryonic fraction ${\Omega}_{\mboxs{B}} = 0.05$ (solid line) and 
${\Omega}_{\mboxs{B}} = 0.1$ (dashed line). Filled squares: numerical
simulations with ${\Omega}_{\mboxs{B}} = 0.05$. Crosses: numerical
simulations with ${\Omega}_{\mboxs{B}} = 0.1$.
}
\end{figure*}

%% file: rad.bbl
\begin{thebibliography}{}
 \bibitem[\protect\citename{Anninos \& Norman }1994]{anno} Anninos,
W.Y., Norman, M.L., 1994, ApJ, 429, 434
1985, ApJSS, 58, 39
 \bibitem[\protect\citename{Babul \& Katz }1993]{bk} Babul, A., Katz,
N., 1993, ApJ, 406, L51
 \bibitem[\protect\citename{Benz }1991]{be1} Benz W., 1991, Late 
stages of stellar evolution. Computational menthods in astrophysical
hydrodynamics, Lecture Notes in Physics 373, ed. C.B. de Loore,
Springer, Berlin
 \bibitem[\protect\citename{Bertschinger }1985]{bert} Bertschinger E.,
1985, ApJSS, 58, 39
 \bibitem[\protect\citename{Bertschinger }1989]{bert2} Bertschinger
E., 1989, ApJ, 318, 66
 \bibitem[\protect\citename{Binney }1977]{bi1} Binney J., 1977, ApJ,
 215, 483
 \bibitem[\protect\citename{Blumenthal \itl{et al.} }1984]{bfpr}
Blumenthal G.R., Faber S.M., Primack J.R., Rees M.J., 1984, Nature,
311, 517
 \bibitem[\protect\citename{Casertano \& Van Gorkom }1991]{cvg} 
Casertano S., van Gorkom J.H., 1991, AJ, 101, 1231
 \bibitem[\protect\citename{Cen }1992]{ce1} Cen R., 1992, ApJSS, 78, 341
 \bibitem[\protect\citename{Cole \itl{et al.} }1994]{cafnz} Cole S.,
Arag\'{o}n-Salamanca A., Frenk C.S., Navarro J.F., Zepf S.E., 1994,
 MNRAS, 271, 781
 \bibitem[\protect\citename{Davis \itl{et al.} }1985]{defw} Davis M.,
Efstathiou G., Frenk C.S., White S.D.M., 1985, ApJ, 292, 371
 \bibitem[\protect\citename{Dekel \& Silk }1986]{ds} Dekel A., Silk
J., 1986, ApJ, 303, 39
 \bibitem[\protect\citename{Efstathiou }1992]{ef} Efstathiou, G.,
 1992, MNRAS, 256, 43P
 \bibitem[\protect\citename{Fillmore \& Goldreich }1984]{fg} Fillmore
J.A., Goldreich P., 1984, ApJ, 281, 1
 \bibitem[\protect\citename{Forcada-Mir\'{o} }1997]{fo1}
 Forcada-Mir\'{o} M.I., 1997, Ph.D. Thesis, University of Cambridge
 \bibitem[\protect\citename{Gaetz, Edgar \& Chevalier }1986]{gec}
Gaetz T.J., Edgar R.J., Chevalier R.A., 1988, ApJ, 329, 927
 \bibitem[\protect\citename{Gott }1975]{go1} Gott III J.R., 1975, ApJ,
201, 296
 \bibitem[\protect\citename{Gunn }1977]{gu1} Gunn J.E., 1977, ApJ, 218, 592
 \bibitem[\protect\citename{Haiman \itl{et al.} }1996]{zol} Haiman,
Z., Thoul, A.A., Loeb, A., 1996, ApJ, 464, 523
 \bibitem[\protect\citename{Hindmarsh }1983]{phi} Hindmarsh A.C.,
1983, Scientific Computing, R.S. Stepleman et al.\ eds.\,
North-Holland, Amsterdam
 \bibitem[\protect\citename{Imamura, Wolf \& Durisen }1984]{iwd}
Imamura J.N., Wolff M.T., Durisen R.H., 1984, ApJ, 276, 667
 \bibitem[\protect\citename{Katz \itl{et al.} }1992]{khw} Katz N.,
 Hernquist L., Weinberg D.H., 1992, ApJ, 399, L109
 \bibitem[\protect\citename{Kauffmann  \itl{et al.} }1993]{kwg}
Kauffmann G., White S.D.M., Guiderdoni B., 1993, MNRAS, 264, 201
 \bibitem[\protect\citename{Kauffmann  \itl{et al.} }1997]{kns}
 Kauffmann G., Nusser, A., Steinmetz, M., 1997, MNRAS, 286, 795
 \bibitem[\protect\citename{Kormendy }1988]{kor}
Kormendy J., 1988, in Origin, Structure, and Evolution of Galaxies,
Fang, L. Z. ed.\ , World Scientific Publishing, Singapore, p.\ 252
 \bibitem[\protect\citename{Langer \itl{et al.} }1982]{lcs} Langer S.H.,
Chanmugam G., Shaviv G., 1982, ApJ, 258, 289
 \bibitem[\protect\citename{Larson }1974]{lar1} Larson D., 1974,
MNRAS, 166, 585
 \bibitem[\protect\citename{Mather \itl{et al.} }1994]{mat}
Mather, J.C. et al., 1994, ApJ, 420, 439
 \bibitem[\protect\citename{Miralda-Escud\'{e} \& Rees }1994]{mer}
Miralda-Escud\'{e} J., Rees M.J., 1994, MNRAS, 266, 343
 \bibitem[\protect\citename{Navarro \& White }1993]{nw} Navarro J.F.,
 White S.D.M., 1993, MNRAS, 265, 271
 \bibitem[\protect\citename{Ostriker }1993]{os1} Ostriker J.P., 1993,
ARA\&A, 31, 689
 \bibitem[\protect\citename{Ostriker \& Steinhardt }1995]{oss} 
Ostriker J.P., Steinhardt P.J., 1995, Nature, 377, 600
 \bibitem[\protect\citename{Peebles }1967]{pe1} Peebles P.J.E., 1967,
ApJ, 147, 859
\bibitem[\protect\citename{Peebles }1993]{pe4} Peebles P.J.E., 1993, 
Principles of Physical Cosmology, Princeton, Princeton 
University Press
\bibitem[\protect\citename{Press \itl{et al.} }1992]{press} Press
W.H., Teukolsky S.A., Vetterling W.T., Flannery B.P., 1992, Numerical
Recipes in Fortran, Cambrdige, Cambridge University Press
 \bibitem[\protect\citename{Rees \& Ostriker }1977]{ro} Rees M.J.,
Ostriker J.P., MNRAS, 179, 541
 \bibitem[\protect\citename{Shapiro, Giroux \& Babul }1994]{sgb} 
Shapiro P.R., Giroux, M.L., Babul, A., 1994, ApJ, 427, 25
 \bibitem[\protect\citename{Shapiro \& Moore }1976]{sham}
Shapiro P.R., Moore, R.T., 1976, ApJ, 207, 460
 \bibitem[\protect\citename{Shapiro \& Struck-Marcell }1984]{sha1}
Shapiro P.R., Struck-Marcell C., 1984, ApJSS, 57, 205
 \bibitem[\protect\citename{Shull \& McKee }1979]{shm} Shull J.M.,
McKee C.F., 1979, ApJ, 227, 131
 \bibitem[\protect\citename{Silk }1977]{si} Silk J., 1977, ApJ, 211, 638
 \bibitem[\protect\citename{Spitzer }1962]{spi} Spitzer L., 1962,
Physics of Fully Ionized Gases, 2nd.\ ed.\, New York, Interscience
 \bibitem[\protect\citename{Steinmetz }1996]{st1} Steinmetz M., 1996,
MNRAS, 278, 1005
 \bibitem[\protect\citename{Strickland \& Blondin }1995]{stb}
Strickland R., Blondin J.M., 1995, ApJ, 449, 727
 \bibitem[\protect\citename{Thoul \& Weinberg }1995]{twe1} Thoul A.A.,
Weinberg D.H., 1995, ApJ, 442, 480
 \bibitem[\protect\citename{Thoul \& Weinberg }1996]{twe2} Thoul A.A.,
Weinberg D.H., 1996, ApJ, 465, 608
 \bibitem[\protect\citename{Tscharnuter \& Winkler }1979]{tsw}
Tscharnuter W.M., Winkler K.H., 1979, CompPhysComm, 18, 171
 \bibitem[\protect\citename{White, Scott \& Silk }1994]{wss} White M.,
Scott D., Silk J., 1994, ARA\&A, 32, 319
 \bibitem[\protect\citename{White \& Frenk }1991]{wf} White S.D.M.,
Frenk C.S., 1991, ApJ, 379, 52
 \bibitem[\protect\citename{White \& Rees }1978]{wr} White, S.D.M.,
 Rees, M.J., 1978, MNRAS, 183, 341
 \bibitem[\protect\citename{White \itl{et al.} }1993]{wn} White S.D.M.,
Navarro J.F., Evrard A.E., Frenk C.S., Nature, 366, 429
 \bibitem[\protect\citename{White \& Zaritsky }1992]{wz} White S.D.M.,
Zaritsky D., ApJ, 394, 1
 \bibitem[\protect\citename{Zaritsky \& White }1994]{zw1} Zaritsky D.,
White S.D.M., 1994, ApJ, 435, 599
\end{thebibliography}
